\documentclass[aps,prb,reprint,showpacs]{revtex4-1}
\usepackage{amsmath}
\usepackage{SIunits}
\usepackage{graphicx}% Include figure files

\begin{document}
%
%
% Define new commands:
%
\renewcommand{\a}[1]{\ensuremath{\hat{a}_{#1}}}
\newcommand{\ac}[0]{\ensuremath{\hat{a}_{\mathrm{c}}}}
\renewcommand{\ap}[0]{\ensuremath{\hat{a}_{\mathrm{p}}}}
\newcommand{\aR}[0]{\ensuremath{\hat{a}_{\mathrm{R}}}}
\newcommand{\aT}[0]{\ensuremath{\hat{a}_{\mathrm{T}}}}
\newcommand{\adag}[1]{\ensuremath{\hat{a}_{#1}^{\dagger}}}
\newcommand{\adagc}[0]{\ensuremath{\hat{a}^{\dagger}_{\mathrm{c}}}}
\newcommand{\adagp}[0]{\ensuremath{\hat{a}^{\dagger}_{\mathrm{p}}}}
\newcommand{\adagR}[0]{\ensuremath{\hat{a}_{\mathrm{R}}^{\dagger}}}
\newcommand{\adagT}[0]{\ensuremath{\hat{a}_{\mathrm{T}}^{\dagger}}}
\newcommand{\betain}[0]{\ensuremath{\beta_{\mathrm{in}}}}
\newcommand{\bra}[1]{\ensuremath{\left<#1\right|}}
\newcommand{\Cm}[0]{\ensuremath{C_{m}}}
\newcommand{\Cmdag}[0]{\ensuremath{C_{m}^{\dagger}}}
\newcommand{\Deltaa}[0]{\ensuremath{\Delta_{\mathrm{a}}}}
\newcommand{\Deltaac}[0]{\ensuremath{\Delta_{\mathrm{ac}}}}
\newcommand{\Deltac}[0]{\ensuremath{\Delta_{\mathrm{c}}}}
\newcommand{\gammapar}[0]{\ensuremath{\gamma_{\parallel}}}
\newcommand{\gammaperp}[0]{\ensuremath{\gamma_{\perp}}}
\newcommand{\gp}[0]{\ensuremath{g_{\mathrm{p}}}}
\renewcommand{\H}[0]{\ensuremath{\hat{H}}}
\newcommand{\kappap}[0]{\ensuremath{\kappa_{\mathrm{p}}}}
\newcommand{\ket}[1]{\ensuremath{\left|#1\right>}}
\renewcommand{\L}[0]{\ensuremath{\mathcal{L}}}
\newcommand{\mean}[1]{\ensuremath{\left<#1\right>}}
\newcommand{\omegaa}[0]{\ensuremath{\omega_{\mathrm{a}}}}
\newcommand{\omegac}[0]{\ensuremath{\omega_{\mathrm{c}}}}
\newcommand{\omegaL}[0]{\ensuremath{\omega_{\mathrm{L}}}}
\newcommand{\pauli}[0]{\ensuremath{\hat{\sigma}}}
\newcommand{\pexc}[0]{\ensuremath{p_{\mathrm{exc}}}}
\newcommand{\taujitter}[0]{\ensuremath{\tau_{\mathrm{jit}}}}
\newcommand{\Tr}[0]{\ensuremath{\mathrm{Tr}}}
\newcommand{\vac}[0]{\ensuremath{\hat{v}}}

\title{Reflectivity and transmissivity of an optical cavity coupled to
  two-level atoms: Coherence properties and the influence of atomic
  phase noise}

\author{B. Julsgaard}
\email{Electronic mail: brianj@phys.au.dk}
\author{K. M{\o}lmer}
\affiliation{Department of Physics and Astronomy, Aarhus University,
  Ny Munkegade 120, DK-8000 Aarhus C, Denmark.}


\date{\today}

\begin{abstract}
  We consider $N$ identical two-level atoms coupled to an optical
  cavity, which is coherently driven by an external field. In the
  limit of small atomic excitation, the reflection and transmission
  coefficients for both fields and intensities are calculated
  analytically. In addition, the frequency content of the cavity field
  and hence also the emission spectrum of the cavity is determined. It
  is discussed in particular how individual collisional dephasing and
  common atomic energy-level fluctuations prevent the cavity field
  from being in a coherent state, which in turn affects the outgoing
  fields.
\end{abstract}


\maketitle

\section{Introduction}
In the field of cavity-quantum-electrodynamics, the coupling between a
single two-level system and the electro-magnetic field is enhanced by
a cavity, which allows for the study of light-matter interactions in
the most fundamental way
\cite{Kimble.PhysicaScripta.T76.127(1998)}. Such a setup has been
implemented in various physical systems, e.g.: A Rydberg atom coupled
to a micro-wave cavity \cite{Brune.PhysRevLett.76.1800(1996)}, an
alkali atom coupled to an optical cavity
\cite{Boca.PhysRevLett.93.233603(2004),
  Maunz.PhysRevLett.94.033002(2005)}, a super-conducting qubit coupled
to a transmission line resonator \cite{Wallraff.Nature.431.162(2004),
  Mallet.NaturePhys.5.791(2009), Wang.PhysRevLett.103.200404(2009)},
and a semi-conductor quantum dot coupled to a photonic-crystal cavity
\cite{Reithmaier.Nature.432.197(2004), Yoshie.Nature.432.200(2004)}.
Ensembles of two-level atoms, each coupled weakly to the cavity field,
are also able to present an effective strong cooperative coupling
\cite{Herskind.NaturePhys.5.494(2009),
  Schuster.PhysRevLett.105.140501(2010),
  Kubo.PhysRevLett.105.140502(2010)}. The light-matter coupling gives
rise to a normal-mode splitting of the resonance frequencies [known as
the vacuum-Rabi splitting as explained by the Jaynes-Cummings model
\cite{Jaynes.ProcIEEE.51.89(1963)} for a single atom], which can be
detected dynamically as oscillations between atomic and optical
excitations \cite{Brune.PhysRevLett.76.1800(1996)}, as a double-peak
in the steady-state fluorescence spectrum
\cite{Sanchez-Mondragon.PhysRevLett.51.550(1983)}, or as a doublet
structure in the cavity-transmission profile
\cite{Zhu.PhysRevLett.64.2499(1990)}. The transmitted and reflected
fields can be exploited in real-time detection and control of trapped
atoms \cite{Hood.Science.287.1447(2000), Pinkse.Nature.404.365(2000)},
and the reflection and transmission profile of these fields has
previously been calculated using both classical
\cite{Zhu.PhysRevLett.64.2499(1990)} and quantum theories
\cite{Albert.Arxiv.1108.0528v1}. While the atomic-dipole-moment decay
was most likely lifetime limited in the traditional experiments with
free atoms in vacuum \cite{Brune.PhysRevLett.76.1800(1996),
  Boca.PhysRevLett.93.233603(2004),
  Maunz.PhysRevLett.94.033002(2005)}, the modern solid-state
implementation has called for an increased attention to dephasing
mechanisms in the case of both fluorescence
\cite{Laucht.PhysRevLett.103.087405(2009)} and transmission
measurements \cite{Srinivasan.Nature.450.862(2007)}. The present work
derives analytical expressions for the transmitted and reflected
intensities, which under the influence of atomic dephasing noise do
not coincide with the modulus square of the transmitted and reflected
fields. In addition, the spectrum of the field emitted from the cavity
is determined.

\section{An empty cavity with phase noise}
\label{sec:empty-cavity}
\begin{figure}
  \centering
  \includegraphics[width=\linewidth]{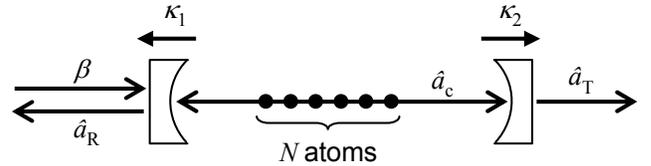}
  \caption{The schematics of the physical system under
    consideration. An optical cavity contains a cavity field, $\ac$,
    which couples equally to $N$ two-level atoms. The field-decay
    rates through the two mirrors are denoted by $\kappa_1$ and
    $\kappa_2$, which connect the cavity field to the coherent input
    field, $\beta$, the reflected field, $\aR$, and the transmitted
    field, $\aT$.}
\label{fig:Cavity}
\end{figure}
Before considering the real problem of a cavity containing $N$ atoms
subjected to various decay processes, we wish to motivate the efforts
by a simpler example: An empty cavity subjected to phase noise and
driven coherently by the field $\betain = \beta e^{-i\omegaL t}$
(Fig.~\ref{fig:Cavity} without atoms, $N = 0$). In the frame rotating
at the driving frequency, $\omegaL$, the Hamiltonian of this simple
system is given by:
\begin{equation}
\label{eq:H_empty}
  \H = \hbar\Deltac\adagc\ac +
     i\hbar\sqrt{2\kappa_1}(\beta\adagc - \beta^*\adagc),
\end{equation}
where $\Deltac = \omegac - \omegaL$ is the detuning from the cavity
resonance frequency, $\omegac$. The input field amplitude, $\beta$, is
normalized such that $|\beta|^2$ is the number of photons incident on
the cavity input mirror per second, and the field-decay rate,
$\kappa_1$, at the input mirror connects the cavity field, $\ac$, to
the incident and reflected fields, $\beta$ and $\aR$, by the
input-output formalism of \citet{Collett.PhysRevA.30.1386(1984)}:
\begin{equation}
\label{eq:Input_output_fields}
  \begin{split}
    \aR &= \sqrt{2\kappa_1}\ac - (\beta+\vac_1), \\
    \aT &= \sqrt{2\kappa_2}\ac - \vac_2.
  \end{split}
\end{equation}
The second line determines the transmitted field, $\aT$, using the
field-decay rate, $\kappa_2$, of the output mirror. In the above
equations $\vac_1$ and $\vac_2$ operate on vacuum states and will have
no further impact on the calculations. In order to account for decay
processes, we take the approach of the master equation,
$\frac{\partial\rho}{\partial t} = \frac{1}{i\hbar}[\H,\rho] +
\L(\rho)$, where the Lindblad part, $\L(\rho)$, is a sum of terms:
\begin{equation}
\label{eq:Standard_Lindblad}
  \L_m(\rho) = -\frac{1}{2}(\Cmdag\Cm\rho + \rho\Cmdag\Cm) + \Cm\rho\Cmdag.
\end{equation}
The leakage of photons from the cavity is modeled by a Lindblad term
with $C_1 = \sqrt{2\kappa}\ac$, where $\kappa = \kappa_1 + \kappa_2$
is the total cavity-field decay rate.

\begin{figure}
  \centering
  \includegraphics[width=\linewidth]{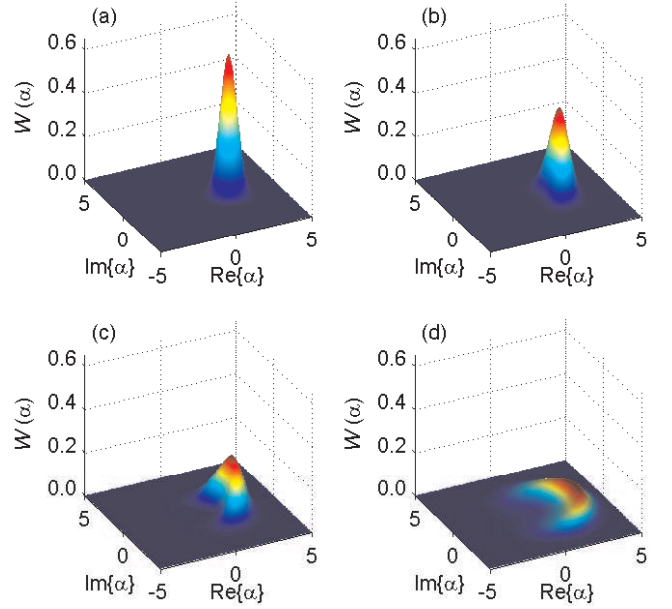}
  \caption{The Wigner function for the cavity field for various
    contributions of phase noise. In all panels, $\mean{\ac} = 2$, and
    the value of $(\kappa\taujitter)^{-1}$ is varied as: (a) $0.0$, (b)
    $0.1$, (c) $0.3$, and (d) $1.0$.}
\label{fig:WignerFunctions}
\end{figure}

Now, we wish to include phase noise in the cavity. Physically, this
could arise from a fast jitter of one of the mirrors such that the
cavity resonance frequency becomes $\omegac' = \omegac +
\varepsilon(t)$, where $\varepsilon(t)$ is a real function fulfilling
$\mean{\varepsilon(t)} = 0$, and $\mean{\varepsilon(t)\varepsilon(t')}
= \frac{2}{\taujitter}\delta(t-t')$. Computationally, the
time-dependent term, $\H_{\mathrm{jit}} =
\hbar\varepsilon(t)\adag{c}\a{c}$, could be added to the Hamiltonian;
however, the fact that the noise is delta-correlated enables a
simpler, time-independent modeling using a Lindblad term with $C_2 =
\sqrt{\frac{2}{\taujitter}}\adagc\ac$, see
App.~\ref{app:Lindblad_fluctuations}. The master equation now leads to
the dynamical equations for the cavity field and photon number:
\begin{align}
\label{eq:ddt_ac_empty}
  \frac{\partial\mean{\ac{}}}{\partial t} &=
     -(\Gamma+i\Deltac)\mean{\ac{}} +\sqrt{2\kappa_1}\beta, \\
  \frac{\partial\mean{\adagc\ac}}{\partial t} &=
   -2\kappa\mean{\adagc\ac} +\sqrt{2\kappa_1}\left(
    \beta\mean{\adagc}+\beta^*\mean{\ac}\right),
\end{align}
where $\Gamma = \frac{1}{\taujitter}+\kappa$. In steady state, the
mean field amplitude and photon number can be written:
\begin{align}
\label{eq:Empty_cavity_ac}
  \mean{\ac} &= \frac{\sqrt{2\kappa_1}\beta}{\Gamma+i\Deltac}, \\
\label{eq:Empty_cavity_ncav}
  \mean{\adagc\ac} &= \left(1+\frac{1}{\kappa\taujitter}\right)|\mean{\ac}|^2.
\end{align}
This result is important for calculating the reflected $\adagR\aR$ or
transmitted $\adagT\aT$ intensity using
Eqs.~(\ref{eq:Input_output_fields}), and it is evident that the
introduction of phase noise ($\taujitter < \infty$) prevents the
cavity-field steady state from being coherent, which would otherwise
lead to the simpler relation, $\mean{\adagc\ac} = |\mean{\ac}|^2$. The
gradual change from a coherent state, when the contribution
$(\kappa\taujitter)^{-1}$ to Eq.~(\ref{eq:Empty_cavity_ncav})
increases from zero, is shown by the Wigner function
\cite{Gardiner.QuantumNoise} in Fig.~\ref{fig:WignerFunctions}. In our
real $N$-atom problem to be discussed in the next section, the atoms
will effectively add phase noise to the cavity field, and, similar to
the present example, the transmitted and reflected intensities cannot
simply be calculated as the modulus square of the field amplitudes. We
note that the impact of phase noise in a cavity on the density matrix
itself has been studied experimentally by
\citet{Wang.PhysRevLett.103.200404(2009)}.

\section{A cavity filled with atoms}
\label{sec:Include_atoms}
Now, consider the entire setup of Fig.~\ref{fig:Cavity} with $N$
identical two-level atoms placed in the coherently driven optical
cavity. In addition to the empty-cavity Hamiltonian of
Eq.~(\ref{eq:H_empty}), we add the contribution from atoms:
\begin{equation}
\label{eq:Atomic_Hamiltonian}
  \H_{\mathrm{atom}} = \frac{\hbar\Deltaa}{2}
      \sum_{j=1}^N\pauli_z^{(j)}
   + \hbar g \sum_{j=1}^N(\pauli_+^{(j)}\ac + \pauli_-^{(j)}\adagc).
\end{equation}
Here $\Deltaa = \omegaa-\omegaL$ is the detuning of the driving
frequency, $\omegaL$, from the atomic resonance frequency,
$\omegaa$. The coupling constant, $g$, is taken to be equal for all
atoms, the $j$'th of which is described by the Pauli operators,
$\pauli_k^{(j)}$ with $k = +,-,z$. We restrict the calculations to
the regime of low excitation probability, $\pexc =
\frac{1}{2}(\mean{\pauli_z^{(j)}}+1) \ll 1$. This allows for the
Holstein-Primakoff approximation, in which each atom is modeled by a
harmonic oscillator:
\begin{equation}
\label{eq:pauli_to_harm_osc}
  \pauli_-^{(j)} \rightarrow \a{j}, \quad
  \pauli_+^{(j)} \rightarrow \adag{j}, \quad
  \pauli_z^{(j)} \rightarrow 2\adag{j}\a{j}-1.
\end{equation}
With this approximation, the entire system Hamiltonian reads:
\begin{equation}
\label{eq:Hamiltonian_entire}
  \begin{split}
    \H &= \hbar{\Deltac}\adagc\ac + \hbar\Deltaa\sum_{j=1}^N\adag{j}\a{j} \\
       & + \hbar g \sum_{j=1}^N(\adag{j}\ac + \a{j}\adagc)
         + i\hbar\sqrt{2\kappa_1}(\beta\adagc - \beta^*\adagc),
  \end{split}
\end{equation}
where the constant-energy shift from the $-1$-term in $\pauli_z^{(j)}$
in (\ref{eq:pauli_to_harm_osc}) has been omitted. We assume that the
phase noise from the jittering mirror (Sec.~\ref{sec:empty-cavity}) is
absent ($\taujitter = \infty$, i.e.~$C_2 = 0$), but the cavity leakage
is retained and various atomic decay processes will be
included. Firstly, for each two-level atom the decay of population
with rate $\gammapar$ is modeled by $C_3 =
\sqrt{\gammapar}\pauli_-^{(j)} \rightarrow \sqrt{\gammapar}\a{j}$,
where the arrow denotes the replacement suggested by
Eq.~(\ref{eq:pauli_to_harm_osc}). Secondly, dephasing of each
individual atomic dipole moment is accounted for by $C_4 =
\frac{1}{\sqrt{2\tau}} \pauli_z^{(j)} \rightarrow
\sqrt{\frac{2}{\tau}}\adag{j}\a{j}$, where $\tau$ is the dephasing
rate\cite{Moelmer.JOptSocAmB.10.524(1993)} [the constant term $-1$ in
(\ref{eq:pauli_to_harm_osc}) has been removed as it makes no
contribution to the Lindblad operator~(\ref{eq:Standard_Lindblad})
when $C_m$ is hermitian].  Note that this Lindblad term can be derived
as the average effect of fast random fluctuations of the resonance
frequency of each atom along the lines of
App.~\ref{app:Lindblad_fluctuations}. Due to the Holstein-Primakoff
approximation, the above Lindblad terms, $C_3$ and $C_4$, for
atomic-population decay and dipole-dephasing correspond exactly to the
energy and phase decay in the harmonic oscillator
\cite{Gardiner.QuantumNoise}. Finally, in addition to individual
atomic decay and dephasing processes, we add the possibility of
dephasing mechanisms common to all atoms. For instance, a fluctuating
magnetic field could vary the energy levels of an ensemble of magnetic
dipoles in a uniform manner, and laser-intensity fluctuations could
cause common energy-level variations of atoms in an optical dipole
trap. This scenario is modeled by an additional fluctuating term in
the Hamiltonian, $\H_1 = \hbar\varepsilon(t)\sum_j\adag{j}\a{j}$,
where we assume the correlation time of fluctuations to be fast
compared to any other time scale: $\mean{\varepsilon(t)} = 0$, and
$\mean{\varepsilon(t)\varepsilon(t')} =
\frac{2}{\tau'}\delta(t-t')$. As explained in
App.~\ref{app:Lindblad_fluctuations}, such a fluctuating term can be
cast into Lindblad form (\ref{eq:Standard_Lindblad}) with $C_5 =
\sqrt{\frac{2}{\tau'}}\sum_j\adag{j}\a{j}$.

Based on the results in Sec.~\ref{sec:empty-cavity}, we expect atomic
phase noise to destroy the coherence of the cavity field. Conversely,
in absence of phase noise ($\tau = \tau' = \infty$) but maintaining
population decay ($\gammapar > 0$), the cavity field is indeed
coherent with $\mean{\adagc\ac} = |\mean{\ac}|^2$ provided that the
Holstein-Primakoff approximation holds. To see this, observe that the
Hamiltonian~(\ref{eq:Hamiltonian_entire}) is on the form $\H =
\sum_{j,k}q_{j,k}\adag{j}\a{k}$, where $j,k$ are indices for atoms,
the cavity, and the external ($c$-number) driving field, and $q_{j,k}$
are time-independent complex numbers. In the Heisenberg picture:
$\frac{\partial}{\partial t}\a{j}(t) = \frac{1}{i\hbar}\sum_k
q_{j,k}\a{k}(t)$, which in turn means that the vector $\hat{\vec{a}}$
of all annihilation operators, $\a{j}$, evolve as $\hat{\vec{a}}(t) =
\exp(-\frac{i}{\hbar}\mathbf{Q}t)\hat{\vec{a}}(0)$, where $\mathbf{Q}$
is the matrix of entries $q_{j,k}$. Now, if at time $t=0$ all
oscillators are in a coherent state with amplitudes $\vec{\alpha}$
(e.g.~all atoms and the cavity in the vacuum state and the external
driving field in a coherent state with amplitude $\beta$) such that
$\hat{\vec{a}}(0)\ket{\psi} = \vec{\alpha}\ket{\psi}$, we find at
later times: $\hat{\vec{a}}(t)\ket{\psi} =
\exp(-\frac{i}{\hbar}\mathbf{Q}t) \hat{\vec{\alpha}}\ket{\psi}$,
i.e.~the state remains an eigenstate for all annihilation operators,
and in particular the steady state of each oscillator must be
coherent. Of course, this observation holds in the Schr\"odinger
picture as well. Now, if population decay (with rate $\gamma_j$) of an
oscillator is introduced, i.e.~the Lindblad term with $C =
\sqrt{\gamma_{j}}\a{j}$ is included in the calculations, the evolution
can be interpreted in the Schr\"odinger picture by the
Monte-Carlo-wave-function approach
\cite{Moelmer.JOptSocAmB.10.524(1993)}. This method states that the
wave function should evolve under the non-Hermitian Hamiltonian, $\H -
\frac{i\hbar}{2}\gamma_j\adag{j}\a{j}$, and occasionally be subjected
to a quantum jump, $\ket{\psi} \rightarrow \a{j}\ket{\psi} =
\alpha_j\ket{\psi}$. The latter clearly preserves $\ket{\psi}$, and
the non-Hermitian term is of the form discussed above and thus also
preserves the coherent state.

\section{Steady-state solutions}

\subsection{Exact expressions}
In order to calculate the relevant physical variables in steady state,
the dynamical equations of the cavity field and atomic dipoles are
deduced as a first step:
\begin{align}
  \frac{\partial\mean{\ac}}{\partial t} &= -(\kappa+i\Deltac)\mean{\ac}
    - ig\sum_{j=1}^N \mean{\a{j}} + \sqrt{2\kappa_1}\beta, \\
  \frac{\partial\mean{\a{j}}}{\partial t} &= -(\gammaperp+i\Deltaa)\mean{\a{j}}
   -ig\mean{\ac},
\end{align}
where $\gammaperp =
\frac{1}{\tau}+\frac{1}{\tau'}+\frac{\gammapar}{2}$. The steady state
for the atomic operators becomes $\mean{\a{j}} =
\frac{-ig\mean{\ac}}{\gammaperp+i\Deltaa}$, which in turn leads to the
cavity-field steady-state value:
\begin{equation}
  \mean{\ac} = \frac{\sqrt{2\kappa_1}\beta}{(\kappa+i\Deltac)(1+v)},
   \quad  v = \frac{g^2N}{(\kappa+i\Deltac)(\gammaperp+i\Deltaa)}.
\end{equation}
The complex field-reflection and transmission coefficients can be
immediately deduced using Eq.~(\ref{eq:Input_output_fields}):
\begin{align}
\label{eq:r_field}
  r = \frac{\mean{\aR}}{\beta}
    &= \frac{2\kappa_1}{(\kappa+i\Deltac)(1+v)} -1 \\
\label{eq:t_field}
  t = \frac{\mean{\aT}}{\beta}
    &= \frac{2\sqrt{\kappa_1\kappa_2}}{(\kappa+i\Deltac)(1+v)}.
\end{align}
These equations incorporate the dephasing mechanism through the
definition of $\gammaperp$ and have been presented previously
\cite{Albert.Arxiv.1108.0528v1,Zhu.PhysRevLett.64.2499(1990)}. However,
a correct treatment of intensities requires the computation of the
number of photons in the cavity. To this end, we deduce the following
equations of motions:
\begin{align}
\notag
  \frac{\partial}{\partial t}\mean{\adagc\ac} = &-2\kappa\mean{\adagc\ac}
    +\sqrt{2\kappa_1}\left[\beta\mean{\adagc}+\beta^*\mean{\ac}\right]
   \label{eq:d_dt_adagc_ac} \\
   &-ig\sum_{j=1}^N\left[\mean{\adagc\a{j}}-\mean{\ac\adag{j}}\right], \\
\notag
  \frac{\partial}{\partial t}\mean{\adagc\a{j}} =
    &-(\kappa+\gammaperp+i\Deltaac)\mean{\adagc\a{j}}
     +\sqrt{2\kappa_1}\beta^*\mean{\a{j}}
    \label{eq:d_dt_adagc_aj}\\
    &-ig\mean{\adagc\ac}+ig\sum_{k=1}^N\mean{\adag{k}\a{j}}, \\
\notag
  \frac{\partial}{\partial t}\mean{\adag{k}\a{j}} =
   & + ig\left[\mean{\adagc\a{j}}-\mean{\ac\adag{k}}\right]
     \label{eq:d_dt_adagck_aj}\\
   &+2\mean{\adag{k}\a{j}}\left[\frac{\delta_{k,j}-1}{\tau}
     -\frac{\gammapar}{2}\right],
\end{align}
where $\Deltaac = \Deltaa-\Deltac$ has been defined. These equations
can be solved in steady state [the details are presented in
App.~\ref{sec:deriv-nc-and-pexc}]:
\begin{align}
\label{eq:ncav_indiv_noise}
  \mean{\adagc\ac} &= \left[1 + \frac{4g^4N(\kappa+\gammaperp)}
    {(\gammaperp^2+\Deltaa^2)\cdot D}\cdot
    \frac{\frac{\gammaperp}{\tau}+\frac{N\gammapar}{2\tau'}}
     {\frac{1}{\tau}+\frac{\gammapar}{2}}\right]|\mean{\ac}|^2, \\
\label{eq:pexc_indiv_noise}
  \pexc &= \frac{2g^2|\mean{\ac}|^2\frac{\gammaperp}{\gammapar}}
     {\gammaperp^2+\Deltaa^2}\left[1-\frac{4\kappa g^2(\kappa+\gammaperp)}
     {\gammaperp\cdot D}\cdot
    \frac{\frac{\gammaperp}{\tau}+\frac{N\gammapar}{2\tau'}}
     {\frac{1}{\tau}+\frac{\gammapar}{2}}\right],
\end{align}
where $D$ is a constant given by:
\begin{equation}
\label{eq:Determinant}
  \begin{split}
  D = &2\kappa\gammapar[(\kappa+\gammaperp)^2+\Deltaac^2] \\
    &+4\kappa g^2(\kappa+\gammaperp)\left[\frac{1+\frac{N}{2}\gammapar\tau}
     {1+\frac{1}{2}\gammapar\tau} + \frac{N\gammapar}{2\kappa}\right].
  \end{split}
\end{equation}
\begin{figure}
  \centering
  \includegraphics[width=\linewidth]{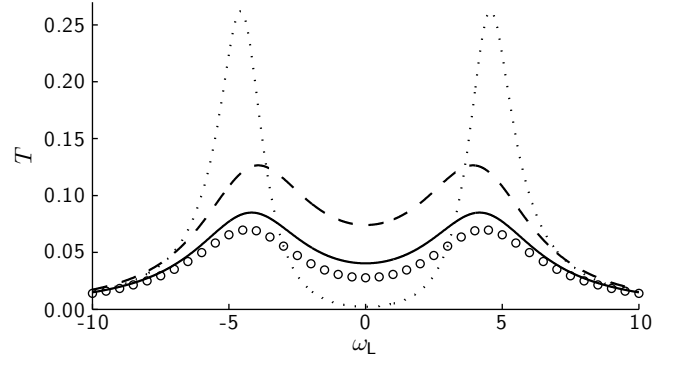}
  \caption{The cavity intensity-transmission coefficient, $T$, as a
    function of driving frequency, $\omegaL$. For all curves, $g = 2$,
    $N = 5$, $2\kappa_1 = 2\kappa_2 = \kappa = 1$, $\omegaa = \omegac
    = 0$, and $\gammapar = 2$. Dotted line: No phase noise
    ($\gammaperp = \frac{\gammapar}{2}$), dashed line: Only common
    phase noise ($\gammaperp = 2\gammapar$), solid line: Only
    individual phase noise ($\gammaperp = 2\gammapar$).  The open
    circles show the modulus square of the field-transmission
    coefficient, $|t|^2$, when $\gammaperp = 2\gammapar$.}
\label{fig:Transmission}
\end{figure}
Equations~(\ref{eq:ncav_indiv_noise}) and~(\ref{eq:pexc_indiv_noise})
present one of the main results of the present paper; they determine
the relevant properties of the coupled atom-cavity system, and with
these at hand the intensity-reflection coefficient, $R =
\mean{\adagR\aR}/|\beta|^2$, and the intensity-transmission
coefficient, $T = \mean{\adagT\aT}/|\beta|^2$, can be stated using
Eq.~(\ref{eq:Input_output_fields}):
\begin{align}
\label{eq:R}
  R &= |r|^2 + \frac{4\kappa_1^2}{\kappa^2+\Deltac^2}\frac{1}{|1+v|^2}\left(
     \frac{\mean{\adagc\ac}}{|\mean{\ac}|^2} -1 \right), \\
\label{eq:T}
  T &= |t|^2\cdot\frac{\mean{\adagc\ac}}{|\mean{\ac}|^2}.
\end{align}
Here $r$ and $t$ are the reflection and transmission coefficients,
respectively, for the fields given by Eqs.~(\ref{eq:r_field})
and~(\ref{eq:t_field}), and the ratio
$\mean{\adagc\ac}/|\mean{\ac}|^2$ is given by
Eq.~(\ref{eq:ncav_indiv_noise}). In the absence of phase noise, this
ratio is unity and the above expressions simplify into $R = |r|^2$ and
$T = |t|^2$.  In order to exemplify the practical impact of phase
noise, consider Fig.~\ref{fig:Transmission}, which shows the
transmission profile for various decay parameters. In comparison to
the case with no phase noise (dotted line), the transmission profile
broadens and decreases in magnitude when phase noise is added (solid
and dashed lines) since $\gammaperp$ increases from
$\frac{\gammapar}{2}$ to $2\gammapar$, which in turn decreases the
magnitude of $|t|^2$ in Eq.~(\ref{eq:T}). However, the transmission
remains higher than $|t|^2$ (open circles) due to the effect of
Eq.~(\ref{eq:ncav_indiv_noise}), and this effect is larger for the
common phase noise (dashed line) as compared to individual case (solid
line). On a relative scale, this effect is most pronounced at $\omegaL
= 0$ [at atomic resonance, $\Deltaa = 0$].
\subsection{Approximate results, limits}
\begin{figure}
  \centering
  \includegraphics[width=\linewidth]{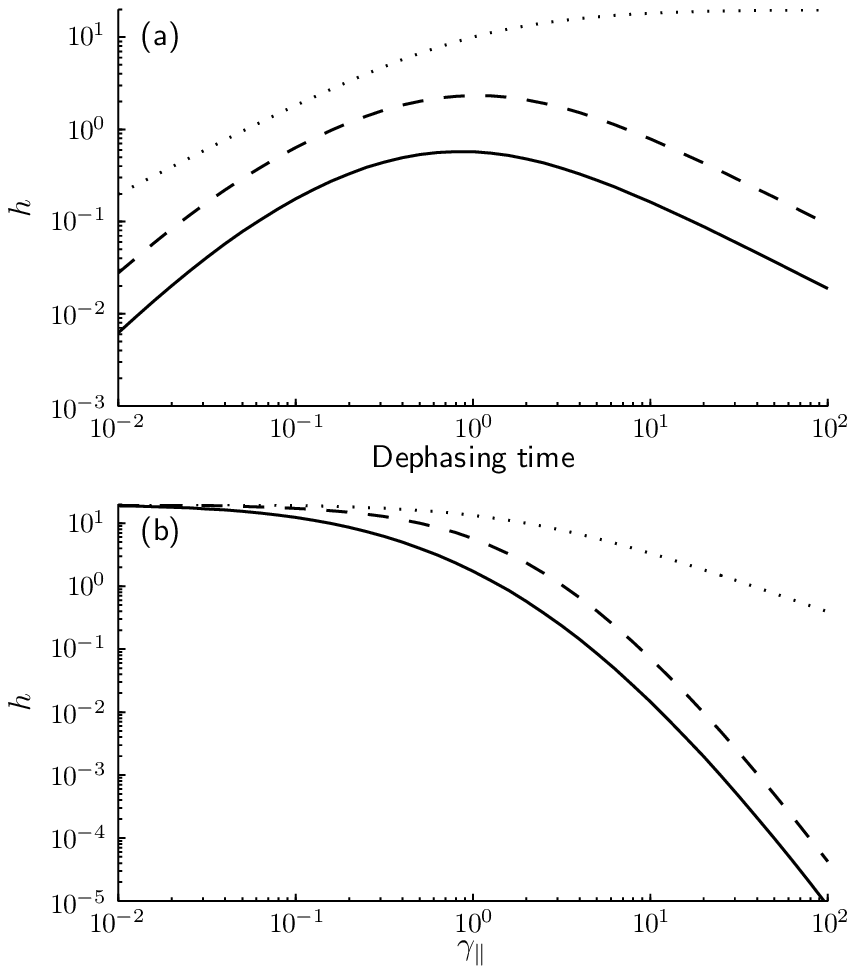}
  \caption{Both panels show on the vertical scale the height, $h$, of
    the Lorentzian contribution: $\mean{\adagc\ac} = [1 + h\cdot
    \frac{\gammaperp^2}{\gammaperp^2+\Deltaa^2}]\cdot|\mean{\ac}|^2$,
    while the atomic decay parameters are varied. Solid line: Only
    individual phase noise ($\tau' = \infty$). Dashed line: Only
    common phase noise ($\tau = \infty$). The dotted line shows the
    cooperativity parameter $C = \frac{g^2N}{\kappa\gammaperp}$. In
    panel (a) the dephasing time ($\tau$ or $\tau'$) is varied with
    $\gammapar = 2$. In panel (b) $\gammapar$ is varied with the
    dephasing time ($\tau$ or $\tau'$) set to unity. For both panels:
    $g = 2$, $N = 5$, $2\kappa_1 = 2\kappa_2 = \kappa = 1$, and
    $\Deltaac = 0$.}
\label{fig:Height}
\end{figure}

Equations~(\ref{eq:ncav_indiv_noise}) and~(\ref{eq:pexc_indiv_noise})
are quite involved, but they reduce to simpler forms in interesting
limiting cases. Consider the fraction (which is common in both
equations):
\begin{equation}
\label{eq:LongFraction}
  \frac{\frac{\gammaperp}{\tau}+\frac{N\gammapar}{2\tau'}}
     {\frac{1}{\tau}+\frac{\gammapar}{2}} = \left\{
       \begin{tabular}{cl}
         $0$                      & when $\tau=\tau'=\infty$,\\
         $\tau^{-1}+(\tau')^{-1}$\: & when $N = 1$, \\
         $\tau^{-1}$               & when $\tau' = \infty$, \\
         $N(\tau')^{-1}$           & when $\tau = \infty$.
       \end{tabular}
     \right.
\end{equation}
This shows, that in absence of phase noise ($\tau = \tau' = \infty$),
the square brackets in Eqs.~(\ref{eq:ncav_indiv_noise})
and~(\ref{eq:pexc_indiv_noise}) reduce to unity. This confirms the
coherent state of the cavity field, $\mean{\adagc\ac} =
|\mean{\ac}|^2$, and Eq.~(\ref{eq:pexc_indiv_noise}) reduces to the
semi-classical expression for the atomic excitation probability in a
classical field [with Rabi frequency, $\chi = 2g\mean{\ac}$]. In the
case of one atom, $N = 1$, there is no difference between common and
individual phase noise, and the two decay channels can be considered
effectively as a single channel with characteristic rate, $\tau^{-1} +
(\tau')^{-1}$.  If only one or the other type of dephasing is present
[individual collision-like dephasing only ($\tau' = \infty$, $\tau <
\infty$) \emph{or} common dephasing only ($\tau = \infty$, $\tau' <
\infty$)], its effect is proportional with the rate of the process,
and for common phase noise it is enhanced by a factor $N$. If we write
Eq.~(\ref{eq:ncav_indiv_noise}) as $\mean{\adagc\ac} = [1 + h\cdot
\frac{\gammaperp^2}{\gammaperp^2+\Deltaa^2}]\cdot|\mean{\ac}|^2$, the
pre-factor $h$ of the Lorentzian function in
Eq.~(\ref{eq:ncav_indiv_noise}) fulfills:
\begin{equation}
\label{eq:limits_on_h}
  h \le \left\{
   \begin{tabular}{cl}
     $C$  & always,\\
     $\frac{2C}{\gammapar}\left(\frac{1}{\tau'}+\frac{1}{N\tau}\right)$
       \:\:  & when $\frac{1}{\tau},\frac{1}{\tau'} \ll \frac{\gammapar}{2}$,
   \end{tabular}
  \right.
\end{equation}
where $C = \frac{g^2N}{\kappa\gammaperp}$ is the cooperativity
parameter. The maximum value $C$ is obtained in the asymptotic limit
$\gammapar \rightarrow 0$. The variations in $h$ as a function of the
atomic decay parameters is illuminated further in
Fig.~\ref{fig:Height}.
In both panels, the cooperativity parameter $C$ (dotted line) clearly
presents an upper limit for $h$. Also, for these examples the common
phase noise (dashed line) gives a larger contribution to $h$ than
individual phase noise (solid line) as indicated by
Eq.~(\ref{eq:limits_on_h}), but otherwise present qualitatively
similar features. In fact, when the decay is lifetime dominated
($\frac{\gammapar}{2} \gg \frac{1}{\tau},\frac{1}{\tau'}$;
right-hand-side limit of the two graphs) the height $h$ decreases as
$(\text{dephasing time})^{-1}$ in panel (a) and as $\gammapar^{-2}$ in
panel (b) as predicted by Eq.~(\ref{eq:limits_on_h}) [note, $C$ scales
as $\gammapar^{-1}$ in this limit]. Conversely, if the dephasing
mechanisms dominate the lifetime decay ($\frac{\gammapar}{2} \ll
\frac{1}{\tau},\frac{1}{\tau'}$; left-hand-side limit of the two
graphs), the cooperativity parameter $C$ is the important figure of
merit. In the limit $\gammapar \rightarrow 0$, one obtains $h
\rightarrow C$ as exemplified in Fig.~\ref{fig:Height}(b). This limit
is reached whenever $\frac{\gammapar}{2} \ll \mathrm{min}\left\{\kappa
  N^{-1}, \: (N\tau)^{-1}, \frac{g^2}{(\kappa+\gammaperp)(1 +
    \frac{\Deltaac^2}{(\kappa+\gammaperp)^2})}\right\}$. For finite
$\gammapar=2$, as exemplified in Fig.~\ref{fig:Height}(a), an
approximate linear scaling of $h$ with $C$ is seen to apply.
\section{The cavity emission spectrum}
\label{sec:Emission_spectrum}
\begin{figure}
  \centering
  \includegraphics[width=\linewidth]{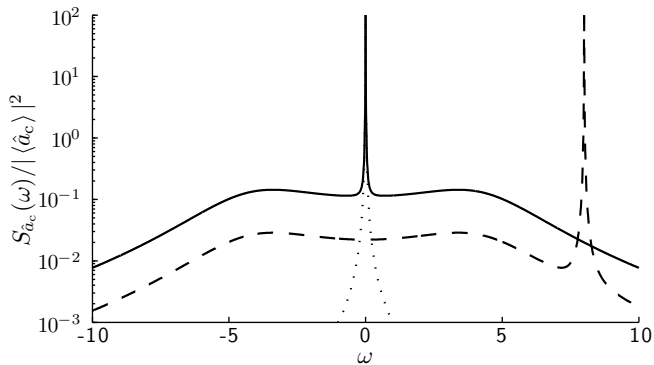}
  \caption{The cavity emission spectrum [note the log scale] modeled
    with a resolution bandwidth of $\pi\kappap = 10^{-2}$. For all
    curves, $g = 2$, $N = 5$, $2\kappa_1 = 2\kappa_2 = \kappa = 1$,
    $\omegaa = \omegac = 0$, and $\gammapar = 2$. Solid and dashed
    lines: Only common phase noise ($\gammaperp = 2\gammapar$) with
    driving frequencies, $\omegaL = 0$ and $\omegaL = 8$,
    respectively. Dotted line: No phase noise, $\omegaL = 0$.}
\label{fig:Spectrum}
\end{figure}

The cavity field, $\ac$, may possess frequency components different
from the driving frequency, $\omegaL$, when phase noise is
present. The spectral density characterizes this effect:
\begin{equation}
  S_{\ac}(\omega) = \frac{1}{2\pi}\int_{-\infty}^{\infty}\negthickspace
  \negthickspace \mean{\adagc(t)\ac(t+\tau)}e^{i\Delta\tau}d\tau,
\end{equation}
where $\Delta = \omega-\omegaL$ accounts for the rotating-frame
picture of $\ac$. Physically, $S_{\ac}(\omega)d\omega$ measures how
much optical energy is present in the cavity within a frequency
bandwidth, $\delta\omega$, around $\omega$. We note that
$\int_{-\infty}^{\infty} \negthickspace S_{\ac}(\omega)d\omega =
\mean{\adagc\ac}$. With the computational details presented in
App.~\ref{app:derive-spectrum}, the spectrum reads for an empty cavity
subjected to phase noise (as described in
Sec.~\ref{sec:empty-cavity}):
\begin{equation}
\label{eq:Spectrum_empty_cavity}
  S_{\ac}(\omega) = \left[
    \frac{\Gamma/\pi}{\Gamma^2+(\omegac-\omega)^2}
    \frac{1}{\kappa\taujitter} + \delta(\omega-\omegaL)\right]|\mean{\ac}|^2,
\end{equation}
where $|\mean{\ac}|^2$ refers to the steady-state cavity field
depending on $\omegaL$ through Eq.~(\ref{eq:Empty_cavity_ac}). The
spectrum is divided into two parts: A broad-band, incoherent
Lorentzian term which is only present when the cavity is subjected to
phase noise ($\taujitter <\infty$), and a coherent part oscillating
exactly at the driving frequency, $\omegaL$. When atoms are included
(as described in Sec.~\ref{sec:Include_atoms}) the spectrum becomes:
\begin{widetext}
\begin{equation}
\label{eq:Spectrum_with_atoms}
  S_{\ac}(\omega) = \left[\frac{(\kappa+\gammaperp)(\kappa\gammaperp+g^2N)+
    \frac{\kappa\gammaperp\Deltaac^2}{\kappa+\gammaperp}}{\pi|g^2N + 
    (\kappa+i[\omegac-\omega])(\gammaperp+i[\omegaa-\omega])|^2} 
  \times \frac{h\gammaperp^2}{\gammaperp^2+(\omegaa-\omegaL)^2} 
  + \delta(\omega-\omegaL)\right]|\mean{\ac}|^2,
\end{equation}  
\end{widetext}
where $h$ is the peak height discussed around
Eq.~(\ref{eq:limits_on_h}). We remind that $h=0$ in absence of phase
noise. Once again, the spectrum is divided into a broad-band,
incoherent part and a delta-function term accounting for the coherent
part. In both Eqs.~(\ref{eq:Spectrum_empty_cavity})
and~(\ref{eq:Spectrum_with_atoms}), the left-most fraction in the
square brackets is a function with unity area being proportional to
$T(\omega) - |t(\omega)|^2$, where $T$ and $t$ are the intensity and
field-transmission profiles, but with the driving frequency,
$\omegaL$, replaced by the observation frequency, $\omega$. These
fractions are then multiplied by the steady-state value of
$\mean{\adagc\ac}-|\mean{\ac}|^2$ (the right-most fractions in the two
equations together with $|\mean{\ac}|^2$), which thus becomes the area
of the incoherent part of the spectrum. Due to the delta-functions,
the coherent part contributes the remaining area, $|\mean{\ac}|^2$,
such that the total area amounts to $\mean{\adagc\ac}$ as expected.
Examples of spectra are presented in Fig.~\ref{fig:Spectrum} with
parameters closely resembling the dashed-line intensity-transmission
profile of Fig.~\ref{fig:Transmission}. As can be seen, it is only
the magnitude of the incoherent spectrum which depends on the driving
frequency, $\omegaL$; its shape corresponds to the difference between
the dashed line and the circles of Fig.~\ref{fig:Transmission}. The
sharp peaks resemble the coherent part, which follows $\omegaL$.
While the coherent part of the reflected and transmitted fields is
related to the cavity field by the input-output relations
Eq.~(\ref{eq:Input_output_fields}), the incoherent, broad-band part of
the spectrum is simply $2\kappa_1$ or $2\kappa_2$ times the incoherent
cavity-field spectrum, $S_{\ac}(\omega)$, for the reflected and
transmitted fields, respectively. We stress that the cavity emission
spectrum [through its exit mirrors] depends on the field-correlation
function, $\mean{\adagc(t)\ac(t+\tau)}$, and does not coincide with
the atomic fluorescence spectrum
\cite{Sanchez-Mondragon.PhysRevLett.51.550(1983)}, which is radiated
into all space and depending on the atomic-dipole correlation
function, $\mean{\pauli_+(t)\pauli_-(t+\tau)}$.

\section{Discussion}
\label{sec:discussion}
The previous sections have identified the practical influence of
atomic phase noise in two cases. Firstly, in a transmission or
reflection measurement where the frequency, $\omegaL$, of the coherent
driving field is varied, the intensity-transmission and reflection
profiles, $T(\omegaL)$ and $R(\omegaL)$, do not coincide with the
modulus square of field-transmission and reflection profiles,
$|t(\omegaL)|^2$ and $|r(\omegaL)|^2$, according to Eqs.~(\ref{eq:R})
and~(\ref{eq:T}). As a result, the pure dephasing-part of the atomic
decay processes can be identified in experiment by e.g.~comparing $T$
and $|t|^2$, since the ratio should present a Lorentzian feature,
$\frac{T}{|t|^2} = 1 +
\frac{h\cdot\gammaperp^2}{\gammaperp^2+\Deltaa^2}$, provided that the
cooperativity parameter is not negligible. The transmitted intensity
can be deduced simply by using a photo-detector, whereas the value of
$|t|^2$ could be determined from measuring two orthogonal quadratures
of the transmitted field in a homodyne detection setup. Secondly, for
a fixed driving frequency, $\omegaL$, the frequency content of the
transmitted field can be determined by connecting the output of a
photo-detector to a spectrum analyzer and searching for the
broad-band, incoherent part as exemplified in Fig.~\ref{fig:Spectrum}.

We remind that all calculations of the present paper are only valid in
the linear regime, i.e.~when atomic excitation is small, which
can be confirmed theoretically by using
Eq.~(\ref{eq:pexc_indiv_noise}), or experimentally by observing that
transmission or reflection properties are independent of the intensity
of the driving field, $\beta$.

\section{Conclusion}
\label{sec:Conclusion}
For an optical cavity coupled to $N$ two-level atoms, the steady-state
reflection and transmission coefficients for both fields and
intensities have been calculated analytically as a function of the
frequency, $\omegaL$, of a coherent driving field, assuming small
atomic saturation. In addition, the frequency spectrum of the field
emitted from the cavity has been determined. It has been demonstrated
that atomic dephasing noise, independently subjected to each atom or
common to all, prevents the cavity field from being in a coherent
state, with two practical implications: (1) intensities do not
coincide with the modulus square of the field amplitudes, and (2) a
broad-band, incoherent part emerges in the cavity-emission spectrum.

\begin{acknowledgments}
  The authors acknowledge support from the EU integrated project AQUTE
  and the EU 7th Framework Programme collaborative project iQIT.
\end{acknowledgments}

\appendix

\section{Transforming a rapidly fluctuating Hamiltonian to a Lindblad
  phase damping term}
\label{app:Lindblad_fluctuations}
This appendix derives the Lindblad form of a phase decay, which is
caused by a fast fluctuating term in the Hamiltonian. Consider the
simple system: $\H = \hbar\varepsilon(t)\hat{O}$, where
$\varepsilon(t)$ is a real function fulfilling: $\mean{\varepsilon(t)}
= 0$ and $\mean{\varepsilon(t)\varepsilon(t')} = D\delta(t-t')$, and
$\hat{O}$ is a time-independent hermitian operator. The time evolution
operator is given by: $\hat{U}(t_1,t_0) =
\exp(-i\hat{O}\int_{t_0}^{t_1}\varepsilon(t)dt)$, which in turn
evolves the density matrix as: $\rho(t_1) = \hat{U}(t_1,t_0) \rho(t_0)
\hat{U}^{\dagger}(t_1,t_0)$. By expanding the time-evolution operator
to second order (which makes sense when $t_1-t_0$ is sufficiently
small):
\begin{equation}
  \hat{U}(t_1,t_0) \approx 1 -i\hat{O}\int_{t_0}^{t_1}
   \negthickspace\negthickspace \varepsilon(t)dt -
    \frac{1}{2}\hat{O}^2
   \int_{t_0}^{t_1} \negthickspace\negthickspace
   \int_{t_0}^{t_1} \negthickspace\negthickspace
     \varepsilon(t)\varepsilon(t')dt dt',
\end{equation}
it is possible to express the density matrix as:
\begin{widetext}
  \begin{equation}
  \begin{split}
  \rho(t_1) &\approx \rho(t_0) -i[\hat{O},\rho(t_0)]\int_{t_0}^{t_1}
   \negthickspace\negthickspace \varepsilon(t)dt
   -\frac{1}{2}\int_{t_0}^{t_1}\negthickspace\negthickspace
   \int_{t_0}^{t_1}\negthickspace\negthickspace
   \varepsilon(t)\varepsilon(t')dt dt'\left[\hat{O}^2\rho(t_0) +
    \rho(t_0)\hat{O}^2 -2\hat{O}\rho(t_0)\hat{O}\right] \\
   &\rightarrow \rho(t_0)
   -\frac{D}{2}(t_1-t_0)\left[\hat{O}^2\rho(t_0) +
    \rho(t_0)\hat{O}^2 -2\hat{O}\rho(t_0)\hat{O}\right],
  \end{split}
  \end{equation}
\end{widetext}
where the arrow denotes averaging over the fast fluctuations of
$\varepsilon(t)$. Now, by taking the limit, $t_1\rightarrow t_0$, we
reach the standard Lindblad form of Eq.~(\ref{eq:Standard_Lindblad})
with $C = \sqrt{D}\hat{O}$.

\section{Derivation of $\mean{\adagc\ac}$ and $\pexc$}
\label{sec:deriv-nc-and-pexc}
The steady-state solution of
Eqs.~(\ref{eq:d_dt_adagc_ac})-(\ref{eq:d_dt_adagck_aj}) is derived in
the following. Firstly, in Eq.~(\ref{eq:d_dt_adagck_aj}) set $k = j$
and perform the summation over $j$. By adding the resulting right-hand
side (which is zero in steady state) to Eq.~(\ref{eq:d_dt_adagc_ac}),
we obtain:
\begin{equation}
\label{eq:Energy_cons}
  2\kappa\mean{\adagc\ac}+N\gammapar\pexc =
  \sqrt{2\kappa_1}\left[\beta\mean{\adagc}+\beta^*\mean{\ac}\right],
\end{equation}
where $\pexc = \frac{1}{N}\sum_j\mean{\adag{j}\a{j}}$ is the
mean atomic excitation. The above equation states
the conservation of energy: On the left-hand side the two terms
describe the energy loss by cavity leakage and atomic population
decay, which must be balanced by the right-hand side --- the energy
delivered by the coherent driving field. Secondly, an explicit
expression for $\sum_{j,k}\mean{\adag{k}\a{j}}$ can be obtained from
Eq.~(\ref{eq:d_dt_adagck_aj}):
\begin{equation}
\label{eq:Mean_of_Sum_adagk_aj}
  \sum_{j,k=1}^N\mean{\adag{k}\a{j}} = N\pexc\frac{1+\frac{N}{2}\gammapar\tau}
    {1+\frac{1}{2}\gammapar\tau}.
\end{equation}
This can be inserted into Eq.~(\ref{eq:d_dt_adagc_aj}) leading to an
expression for $\sum_j\mean{\adagc\a{j}}$, which again is inserted
into Eq.~(\ref{eq:d_dt_adagck_aj}) [being summed over $j=k$]. As a
result, another equation involving $\mean{\adagc\ac}$, $\pexc$, and
the already known cavity field $\mean{\ac}$, is obtained, and we are
left with two equations involving two unknowns:
\begin{equation}
  \begin{bmatrix}
    a & b \\ c & d
  \end{bmatrix}
  \begin{bmatrix}
    \pexc \\ \mean{\adagc\ac}
  \end{bmatrix}
  =
  \begin{bmatrix}
    e \\ f
  \end{bmatrix}
  |\mean{\ac}|^2,
\end{equation}
where the constants $a$, $b$, $c$, $d$, $e$, and $f$ are given by:
\begin{equation}
  \begin{split}
  a &= \gammapar[(\kappa+\gammaperp)^2+\Deltaac^2]+2g^2(\kappa+\gammaperp)
       \frac{1+\frac{N}{2}\gammapar\tau}{1+\frac{1}{2}\gammapar\tau}, \\
  b & = -2g^2(\kappa+\gammaperp), \\
  c &= N\gammapar, \\
  d &= 2\kappa, \\
  e &= \frac{2g^2}{\gammaperp^2+\Deltaa^2}\{
     g^2N(\kappa+\gammaperp)+\kappa(\gammaperp^2-\Deltaa^2) \\
     &\qquad\qquad\qquad
     +\gammaperp(\kappa^2+\Deltac^2)-2\gammaperp\Deltaa\Deltac\}, \\
  f &= 2\kappa+\frac{2\gammaperp g^2N}{\gammaperp^2+\Deltaa^2}.
  \end{split}
\end{equation}
The determinant, $D = ad-bc$, of this set of equations was stated in
Eq.~(\ref{eq:Determinant}). Now, $\mean{\adagc\ac}$ and $\pexc$ can be
deduced using Kramer's rule leading to the results of
Eqs.~(\ref{eq:ncav_indiv_noise}) and~(\ref{eq:pexc_indiv_noise}).

\section{Derivation of the cavity-field frequency spectrum}
\label{app:derive-spectrum}
We present here the computational details for the cavity-field
frequency spectrum, $S_{\ac}(\omega)$. Since it is a measure of the
frequency distribution of the optical energy, an actual measurement of
this entity can be modeled by coupling (weakly) an auxiliary
narrow-band probe cavity to the optical cavity and computing the
steady-state photon number of the probe cavity. In other words, the
physical system described by the Hamiltonian, $\H$, is extended to:
\begin{equation}
  \H' = \H + \hbar\Delta\adagp\ap + \hbar\gp(\adagc\ap + \ac\adagp),
\end{equation}
where $\Delta = \omega - \omegaL$ denotes the observation frequency in
the frame rotating at $\omegaL$ and $\gp$ measures the strength of the
coupling. In addition, the (half-width-half-maximum) bandwidth,
$\kappap$, of the probe cavity is modeled by an extra Lindblad term
with $C = \sqrt{2\kappap}\ap$. In steady state we derive:
\begin{equation}
\label{eq:Mean_ap_general}
    \mean{\ap} = \frac{-i\gp\mean{\ac}}{\kappap+i\Delta},
\end{equation}
which must be kept small for maintaining a negligible disturbance of
the cavity field, $\ac$. By defining $\gp \equiv \epsilon\kappap$, the
probe field is at least a factor of $\epsilon$ weaker than $\ac$ for
any value of $\kappap$ and $\Delta$ and it suffices to calculate
$\mean{\adagp\ap}$ to second order in $\epsilon$. Also, for a
sufficiently small $\epsilon$, the probe field $\ap$ can be truncated
to the two lowest Fock states $\{\ket{0}_{\mathrm{p}},
\ket{1}_{\mathrm{p}}\}$ with $\ap =
\ket{0}_{\mathrm{p}}\!\bra{1}_{\mathrm{p}}$, etc. The Hamiltonian
$\H'$ and the entire density matrix $\rho$ can then be expressed as:
\begin{equation}
    \H' =
  \begin{bmatrix}
   \H^{00} & \H^{01} \\ \H^{10} & \H^{11}
  \end{bmatrix}
  =
  \begin{bmatrix}
    \H & \hbar\gp\adagc \\ \hbar\gp\ac & \H + \hbar\Delta\boldsymbol 1
  \end{bmatrix},
\end{equation}
and
\begin{equation}
  \rho =
  \begin{bmatrix}
    \rho^{00} & \rho^{01} \\ \rho^{10} & \rho^{11}
  \end{bmatrix}.
\end{equation}
It follows that $\mean{\adagp\ap} = \Tr(\rho^{11})$ and $\mean{\ap} =
\Tr(\rho^{10})$. Each of the four sub-blocks in $\H'$ have the
dimensionality of the original, unperturbed system, $\H$, and the
evolution can be written:
\begin{align}
\notag
  \frac{\partial\rho^{00}}{\partial t} = 
    &\frac{1}{i\hbar}([\H^{00}\!,\rho^{00}] 
       + \H^{01}\!\rho^{10} - \rho^{01}\!\H^{10}) \\
\label{eq:Probe_cav_general_MasterEq00}
    &+ \L(\rho^{00})
       + 2\kappap\rho^{11}, \\
\notag
  \frac{\partial\rho^{10}}{\partial t} = 
    &\frac{1}{i\hbar}(\H^{10}\!\rho^{00} + 
       \H^{11}\!\rho^{10} - \rho^{10}\!\H^{00} -\rho^{11}\!\H^{10}) \\ 
\label{eq:Probe_cav_general_MasterEq10}
    &+ \L(\rho^{10}) - \kappap\rho^{10},
\end{align}
where $\L(\cdot)$ is the Lindblad super-operator of the original system and
the probe-cavity decay is added separately. Now, in steady state set
Eq.~(\ref{eq:Probe_cav_general_MasterEq00}) equal to zero and compute
the trace:
\begin{equation}
\label{eq:Formula_adagp_ap}
  \mean{\adagp\ap} = \Tr(\rho^{11}) = \frac{i\epsilon}{2}\left[
   \Tr(\adagc\rho^{10})-\Tr(\ac\rho^{01})\right].
\end{equation}
The value of $\Tr(\adagc\rho^{10})$ and its complex conjugate is
required and may be deduced using
Eq.~(\ref{eq:Probe_cav_general_MasterEq10}) [in steady state, multiply
from the left by $\adagc$ and compute the trace]:
\begin{gather}
\notag
  \frac{1}{i\hbar}\Tr([\adagc,\H]\rho^{10}) + \Tr(\adagc\L(\rho^{10})) \\
\label{eq:Probe_cav_general_regressed_eq}
  \qquad\qquad = (\kappap+i\Delta)\Tr(\adagc\rho^{10}) + i\gp\mean{\adagc\ac}.
\end{gather}
This equation has a great similarity with the steady-state equation
for $\mean{\adagc}$. However, we must replace $\mean{\adagc}
\rightarrow \Tr(\adagc\rho^{10})$, add an additional decay channel
with rate $\kappap$, introduce another detuning of $\Delta$, and the
extra term $i\gp\mean{\adagc\ac}$ must be included. The progress from
here depends on the actual physical system as defined by $\H$ and
$\L$, but in practice the above-mentioned similarities allows for
re-using existing equations in the spirit of the quantum regression
theorem.

Now, in order to keep the derivations simple we adopt the physical
system of Sec.~\ref{sec:empty-cavity}, i.e.~an empty, coherently
driven cavity subjected to phase noise. In similarity with
Eq.~(\ref{eq:ddt_ac_empty}), we obtain from
Eq.~(\ref{eq:Probe_cav_general_regressed_eq}):
\begin{equation}
  \begin{split}
    \Tr(\adagc\rho^{10}) &=  
     \frac{-i\gp\mean{\adagc\ac} + \sqrt{2\kappa_1}\beta^*\Tr(\rho^{10})}
     {\kappap+\Gamma-i(\Deltac-\Delta)} \\
   &= \frac{-i\epsilon\kappap\left[\mean{\adagc\ac} + 
    \frac{\Gamma-i\Deltac}{\kappap+i\Delta}|\mean{\ac}|^2\right]}
     {\kappap+\Gamma-i(\Deltac-\Delta)},
  \end{split}
\end{equation}
where Eqs.~(\ref{eq:Empty_cavity_ac}) and~(\ref{eq:Mean_ap_general})
were used in the second step --- note we only require
$\Tr(\adagc\rho^{10})$ to first order in $\epsilon$, and hence the
unperturbed values of $|\mean{\ac}|^2$ and $\mean{\adagc\ac}$ and also
the relation~(\ref{eq:Empty_cavity_ncav}) can be used. Now, insert the
above result into Eq.~(\ref{eq:Formula_adagp_ap}) and re-arrange the
terms to obtain:
\begin{equation}
  \frac{\mean{\adagp\ap}}{\pi\kappap\epsilon^2} = \left[
    \frac{(\kappap+\Gamma)/(\pi\kappa\taujitter)}
   {(\kappap+\Gamma)^2+(\Deltac-\Delta)^2}
    + \frac{\kappap/\pi}{\kappap^2+\Delta^2}\right]
    |\mean{\ac}|^2.
\end{equation}
The Lorentzian intensity-transmission profile of the probe cavity has
an effective width [area-to-height ratio] of
$\pi\kappap$. Furthermore, since $\epsilon^2$ connects the magnitudes
of $|\mean{\ac}|^2$ and $|\mean{\ap}|^2$ through
Eq.~(\ref{eq:Mean_ap_general}), the left-hand side of the above
expression can be interpreted as the amount of energy per frequency
interval in the probe cavity when taking the limit $\kappap
\rightarrow 0$, which leads to
Eq.~(\ref{eq:Spectrum_empty_cavity}). The case of atoms
[Eq.~(\ref{eq:Spectrum_with_atoms})] can be derived in a similar
fashion.

% Uncomment below to make bibliography. 
% \bibliography{bibfile}

% For submitted version, paste below the bbl-file.

\end{document}